\newcommand{\beq}{\begin{equation}}
\newcommand{\eeq}{\end{equation}}
\newcommand{\beqa}{\begin{eqnarray}}
\newcommand{\eeqa}{\end{eqnarray}}

\renewcommand{\lambda}{\ell}

\newcommand{\bk}{{\bf k}}
\newcommand{\br}{{\bf r}}

\documentstyle[aps,prl,twocolumn,epsf]{revtex}

\begin{document}
\pagestyle{arabic}

\twocolumn[\hsize\textwidth\columnwidth\hsize\csname @twocolumnfalse\endcsname


\title{Magnetic impurity  induced $id_{xy}$ component and spontaneous time reversal and parity breaking  in a
$d_{x^2-y^2}$-wave superconductor}

\author{
  A.V.~Balatsky  
}
\address{Theoretical Division, Los Alamos National Laboratory, Los Alamos, 
New Mexico 87545}
  
\date{\today}
\maketitle

\begin{abstract}

It is shown that the patches of {\em complex} $d_{xy}$ component are
generated around magnetic impurity in the presence of the coupling
between orbital moment of the condensate and impurity spin $S_z$.
Locally induced $d_{xy}$ gap leads to the fully gapped quasiparticle
spectrum near impurity.  The phase of the induced $\pm i d_{xy}$
component is determined by impurity spin and averages to zero at high
temperatures. It is suggested that at low temperature the well defined
patches of $d_{xy}$ are formed and they can undergo a phase transition
into phase locked state via Josephson effect.  Violation of
time-reversal symmetry and parity occurs spontaneously via the second
order transition.  In the ordered phase both impurity magnetization
and $d_{xy}$ component of the order parameter develop and are
proportional to each other.

\


\end{abstract}

\

\

]


It is well known that   magnetic impurities destroy the 
singlet superconducting 
state due to spin scattering which breaks pair singlets 
\cite{AG}. In the case 
of the gapless (with the nodes of the gap) d-wave
 superconductor both magnetic 
and nonmagnetic impurities produce a finite density 
of states at zero energy. 
These effects are a simple and direct consequence
 of the lifetime effects 
produced by impurities. These are well
 known ``incoherent'' effects of impurities 
in unconventional superconductors. After recent
experiments   by Movshovich et.al. 
\cite{Roman1}
 we are led to believe that another phenomenon is possible, namely the
transition to the second superconducting phase as a result of
condensate interactions with magnetic impurities. The time reversal
violating state is formed at low energy and the order parameter of the
new phase is $d_{x^2-y^2} + id_{xy}$ ($d+id$). In this phase the
impurity spins acquire nonzero spin density along z-axis, i.e.  out of
plane.

The physical origin of the instability comes from the fact that the
$d+id$ state has an orbital moment which couples to the magnetic
impurity spins. The relevant interaction is the $\hat{L}_zS_z$
coupling between impurity spin $S_z$ and the conduction electron
orbital moment $L_z$:
\beqa
H_{int} = g\sum_i \int d^2r{S_z(\br_i)\over{|\br - \br_i|^3}}\psi^{\dag}_{\br \sigma}[\br \times i\partial_{\br}]_z \psi_{\br \sigma}
\label{g}
\eeqa
where $g$ -- is the coupling constant, $\psi_{\br \sigma}$ -- the
electron annihilation operator and summation is over impurity sites
$i$.  In the pure phase one can think of d-wave state as an equal
admixture of the orbital moment $L_z = \pm 2$ pairs:
\beqa
\Delta_0(\Theta) = \Delta_0 \cos 2\Theta ={\Delta_0\over{2}}(\exp(2i\Theta) + 
\exp(-2i\Theta))
\eeqa
Here $\Theta$ is the 2D planar angle of the momentum on the Fermi surface, 
$\Delta_0$ is the magnitude of the $d_{x^2-y^2}$ component. We consider 2D 
$d_{x^2-y^2}$ superconductor, motivated by the layered   structure of the 
cuprates. In the presence of the (ferromagnetically) ordered impurity spins 
$S_z$ the coefficients of the $L_z = \pm 2$ components will shift {\em linearly} in $S_z$ with {\em opposite} signs:
\beqa
\Delta_0(\Theta) \rightarrow {\Delta_0\over{2}}((1 + g S_z)\exp(2i\Theta)\nonumber\\ 
+ (1 - g S_z)\exp(-2i\Theta))
 = \Delta_0(\Theta) + i \ S_z\Delta_1(\Theta)
\eeqa
 where $\Delta_1(\Theta)  \propto  g/2 \  \sin 2\Theta$ -- is the $d_{xy}$ component. The  relative phase $\pi/2$ of these two order 
parameters comes out naturally  because  $d+id$ state has a noncompensated 
orbital moment $L_z = +2$.

Here I argue that time reversal (T) and parity (P) symmetries can be
 broken {\em spontaneously} in the bulk of the d-wave state due to
 coupling to the impurity spins. Original $d_{x^2-y^2}$ is unstable
 towards the formation of the bulk $d_{x^2-y^2}+id_{xy}$ phase. In the
 new phase both the spontaneous magnetization of impurity spins and
 second component of the order parameter are developed simultaneously.
 To show how complex $d_{xy}$ component appears, I first consider the
 single magnetic impurity and find that the spin-orbit interaction
 between impurity spin and orbital moment of the conduction electron
 generate a finite {\em complex} $d_{xy}$ anomalous amplitude near
 impurity in $d_{x^2-y^2}$ state. This patch of $d_{xy}$ state is
 formed near impurity site, as long as $d_{x^2-y^2}$ amplitude is
 finite, and has a spatial extend of coherence length $\xi_0 =
 20\AA$. It is therefore possible for these patches to form a long
 range phase coherent state at some lower temperature as a result of
 Josephson tunneling between different patches.  I also present a
 macroscopic Ginzburg-Landau functional (GL) and find that there is a
 linear coupling between the original $d_{x^2-y^2}$ order parameter
 $\Delta_0(\Theta) = \Delta_0 \cos 2\Theta$ and the spontaneously
 induced $d_{xy}$ component: $\Delta_1(\Theta) = \Delta_1 \sin
 2\Theta$. The GL functional contains the {\em linear } coupling term:
\beqa
F_{int} =  -{b\over{2i}}(\Delta_0^*\Delta_1 - h.c.)S_z
\label{Fint}
\eeqa
where $b \propto n_{imp} g$ is the macroscopic coupling constant,
$n_{imp}$ is the impurity concentration per unit cell of linear size
$a$, $\Delta_{0,1}, g/a$ have dimension of energy. The time reversal
violation is natural in this case as it allows the order parameter
$\Delta_0 + i\Delta_1$ to couple directly to the impurity spin. This
coupling is possible only for $d+id$ and not for $d+is$ symmetry of
the order parameter. From the GL description it follows that
instability develops as a {\em second order} phase transition where
both the out-of-plane magnetization $S_z$ and $d_{xy}$ component
developed together and are proportional to each other \cite{SNSBal}.

Recent experimental observation of the surface-induced time-reversal
violating state in YBCO suggests that the secondary component of the
order parameter (d+is) can be induced \cite{Green1}. Theoretical
explanation, based on surface-induced Andreev states has been
suggested by Sauls and co-workers \cite{Sauls1}. The source of the
secondary component is the bending of the original $d_{x^2-y^2}$ order
parameter at the surface.

In a different approach Laughlin \cite{Laugh1} argued that the
 $d_{x^2-y^2}$ state is unstable towards $d+id$ state in the bulk in
 the perpendicular magnetic field at low enough temperatures. The time
 reversal and parity are broken by external field in this case. The
 linear coupling of the secondary order parameter to the external
 field is central to his consideration and results in the first order
 phase transition into $d+id$ state. This transition was suggested to
 be responsible for the kink-like feature in the thermal conductivity
 in experiments by Krishana et.al. \cite{Ong1}

Recent experiments reported the anomaly in the thermal conductivity in
Bi2212 at low temperatures: the thermal conductivity of the Bi2212
with Ni impurities was observed to have a sharp reduction at $T^*_c =
200 mK$ \cite{Roman1}. These experimental data indicate the possible
superconducting phase transition in the Bi2212 in the presence of the
magnetic impurities in a certain concentration range. So far the
transition has been seen only in the samples with magnetic impurities,
e.g.  Ni as opposed to the nonmagnetic impurities such as Zn
\cite{Roman1}.  It was reported that the feature in the thermal
conductivity is completely suppressed by applying the field of $H \sim
200 Gauss$.  The low field and the fact that feature disappears is
consistent with the superconducting transition into second
phase. Results presented here might be relevant for the experimentally
observed transition at $T^*_c$ in Bi2212 with Ni.

1. {\bf Single magnetic impurity and $d_{xy}$ patch.}

I begin by considering one impurity at site $\br_i =0$ and interacting
with conduction electrons via $H_{int}$ in Eq.(\ref{g}). Similar to
the approach of \cite{YSR}, one can find the anomalous propagator in
the presence of the single impurity scattering potential:
$F_{\omega_n}(\bk,\bk') = F^0_{\omega_n}(\bk)\delta(\bk-\bk') +
F^1_{\omega_n}(\bk,\bk')$, where $F^0 = {\Delta_0 \cos
2\Theta\over{\omega_n^2 + \xi_{\bk} + \Delta_0^2 \cos^2 2 \Theta}},
G^0 = -{i\omega_n + \xi_{\bk}\over{\omega_n^2 + \xi_{\bk} + \Delta_0^2
\cos^2 2 \Theta}}$ are the pure system propagators,
$F^1_{\omega_n}(\bk,\bk')$ is the correction due to impurity
scattering, $\bk = (k, \Theta)$ are the magnitude and angle of the
momentum $\bk$ on the cylindrical Fermi surface, $\omega_n$ is
Matsubara frequency and $\xi_{\bk} = \epsilon_{\bk}-\mu$ is the
quasiparticle energy, counted form the Fermi surface.  We take $S$ to
be a classical variable and ignore spin flip scattering.  To linear
order in small $g$ one finds :
\beqa
F^1_{\omega_n}(\bk,\bk') =  -i2\pi g S_z G^0_{\omega_n}(\bk) F^0_{\omega_n}(\bk') {[\bk \times \bk']_z\over{|\bk-\bk'|}}
\label{F1}
\eeqa
Where $ F^1_{\omega_n}(\bk,\bk') $ is the function of incoming and
outgoing momenta because of broken translational symmetry. Upon
integrating $F^1$ over $\bk'$ and going to integrated over $\xi_{\bk}$
propagator one finds:
\beqa
F^1_{\omega_n}(\Theta)=\int N_0 d\xi_{\bk} F^1_{\omega_n}(\bk)= i   \Lambda_{\omega_n}(N_0 gS_z)(N_0 \Delta_0)\sin 2 \Theta
\label{F1int}
\eeqa
Here $\Lambda_{\omega_n} \simeq k_F {\pi^2\over{2\sqrt{2}}} \ln
({W\over{\sqrt{\omega_n^2 + \Delta^2_0}}})\langle 1/\sqrt{\omega_n^2 +
\Delta^2_0 \cos^2 2\Theta}\rangle_{\Theta}$ is the model dependent
coupling, $W$ is the energy cutoff, $N_0$ is the Density of States at
the Fermi surface and $\langle \rangle_{\Theta}$ stands for Fermi
surface averaging.
 
The angle dependence of $ F^1_{\omega_n}(\Theta) \sim i gS_z\sin 2
 \Theta \propto k_xk_y$ is the one of $d_{xy}$. Together with the fact
 that this amplitude is complex it indicates the existence of the
 $id_{xy}$ component in the vicinity of magnetic impurity.  This
 result also shows that incoming $d_{x^2-y^2}$-wave state electrons
 have a finite amplitude, linear in $gS_z$, to be scattered into the
 $d_{xy}$ outgoing state via the $\hat{L}_z S_z$ coupling.  From the
 solution Eq.(\ref{F1}) it is easy to see that the typical size of the
 patch, ignoring nodal directions, is given by superconducting
 coherence length $\xi_0 = 20 \AA$. For the relevant concentration of
 Ni $n_{imp} \sim 1\%$ the Ni-Ni distance is about $35 \AA$. The
 patches are thus well overlapping in this limit making phase ordering
 due to tunneling form patch to patch possible. These patches work as
 a microscopic seed of $d_{xy}$ component which grows into true long
 range state at low temperatures $T \leq T^*_c$. Similar result for
 the $d_{xy}$ patches in the mixed state of pure Bi2212 was shown in
 \cite{Rama}, where the role of impurities is assumed by vortices.

Important observable consequence of the {\em local} $d_{xy}$ component
near impurity is that the gap, as seen in STM tunneling near Ni
impurity, will increase and the low energy part of Density of States
will be suppressed because of finite gap everywhere on the Fermi
surface, as opposed to nodes for pure $d_{x^2-y^2}$ state. The
increase of the gap in the Ni-doped Bi2212, compared to the pure case,
was observed in STM tunneling \cite{Han}

Next, I consider simple example of magnetically ordered state: the
ferromagnetically ordered impurity spins. This does not have to be the
case experimentally but makes the point clearer. Calculation, similar
to the one above, yields:
\beqa
F^1_{\omega_n}(\bk) =  -i n_{imp} {2\pi\over{a}} g S_z G^0_{\omega_n}(\bk) [\partial_{\bk} \times \bk]_z F^0_{\omega_n}(\bk)  
\label{F2}
\eeqa
The existence of the homogeneous $d_{xy}$ component
$F^1_{\omega_n}(\bk) \sim -i n_{imp} {2\pi\over{a}} g S_z \Delta_0
\sin 2\Theta$ is evident from this equation. The relative phase of
$d_{xy}$ component with respect to $d_{x^2-y^2}$ is determined by the
sign of $S_z$. The gap $\Delta_1$ has to be determined
seflconsistently provided there is interaction in $xy$ channel. This
interaction does not have to be attractive, since $d_{x^2-y^2}$ plays
the role of the source and $\Delta_1$ will be generated for any sign
of interaction.  I will assume there is such interaction and results
below will be expressed in terms of the induced gap $\Delta_1$.  With
the help of this equation I find for the free energy change due to
$\Delta_1: \ \delta F = 1/2 T\sum_{\omega_n,\bk} F^1_{\omega_n}(\bk)
\Delta^*_1(\bk) + h.c.$
\beqa
\delta F = -i/2 (N_0\Delta^*_1)(N_0\Delta_0) {2\pi\over{a}}(g  S_z)  n_{imp} + h.c. 
\label{dF}
\eeqa
Eq.(\ref{dF}) together with single impurity result Eq.(\ref{F1int})
are the main results of this section. From this equation I find the
linear term Eq.(\ref{Fint}) with
\beqa
b =  N^2_0 g {2\pi\over{a}} n_{imp}
\label{b}
\eeqa

\begin{figure}
\epsfxsize=3.2in
\centerline{\epsfbox{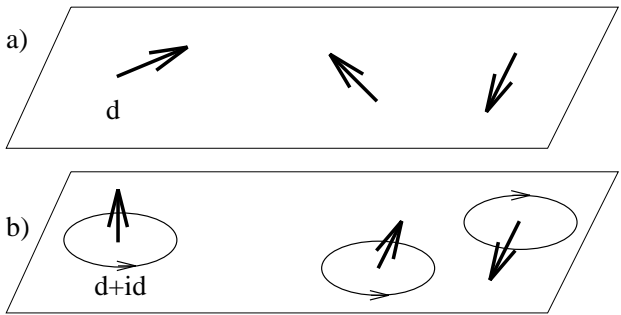}}
\vskip0.25in
 \caption[]{a) Impurity sites with random spins at high temperatures
are shown. Spin flips lead to $d_{xy}$ component averaged to zero. b)
Upon slowing down and freezing of impurity spins the patches of $d \pm
id$ with $L_z = \pm 2$ state near each Ni site are formed, shown with
right(left) circulating current near each impurity site. At low
temperatures Josephson tunneling locks the phase between patches,
leading to the global $d+id$ state.}
\label{patches}
\end{figure}

So far the spin flips were ignored. The relative phase of second
component is determined by the sign of $S_z$. At high temperatures $T
\gg T^*_c$ , when spins are strongly fluctuating, the relative phase
of the $d_{xy}$ component is fluctuating strongly as well. This
phenomenon is an interesting new realization of the superconducting
phase ($d_{xy}$) coupled to the heat bath (fluctuating spins).

If and when the impurity spins are slowing down or even are freezing
 out then the phase scattering time becomes large and the phase
 ordering of the patches is possible, see Fig.1. Measurements indicate
 that spin flips of Ni spins are slowing down at low temperatures
 $T\leq 2 K$.  Specific heat measurements on Ni- doped Bi2212 indicate
 additional entropy, compared to undoped Bi2212, on the order of
 $n_{imp}R\log 3$, accumulated near $1K$ \cite{SNSRoman}.  This
 additional specific heat has a broad maximum around $1K$ and linear
 slope at lower temperatures.  The general shape of the specific heat,
 associated with the impurity spins is strikingly similar to the
 specific heat, observed in spin glasses \cite{SG}. Broad peak in the
 specific heat might indicate the glassy behavior of spins at lower
 temperatures.

2. {\bf Mean field formulation.}

Below I will ignore the fluctuations in the magnetic subsystem and
consider simple mean field theory of the coupled magnetic impurities
and superconducting condensate.  From the specific heat measurements
we know that some type of order (spin-glass or other) might occur at
$T_m \sim 1K$. I will assume that impurities develop a ferromagnetic
order at some temperature $T_m$.  This is a drastic oversimplification
because of the possible spin-glass ordering discussed
above. Nevertheless the model presented below is useful to understand
how the coupling between impurity spins and condensate leads to the
$d+id$ instability of the original state.

I will consider the GL theory of the secondary superconducting
transition: $\Delta_0 \rightarrow \Delta_0 +
\Delta_1$ at $T^*_c$, where both $\Delta_{0,1}$ are homogeneous variables corresponding to the macroscopic ordering. The relative phase of  $\Delta_1$ with respect to the
phase of $\Delta_0$ {\em is not fixed} and will be determined by the
free energy minimization.  Assume that the second transition, if at
all, occurs at $T^*_c \ll T_c$, where $T_c \sim 90 K$ is the first
transition temperature. Hence the order parameter $\Delta_0$, which
can be assumed to be real, is robust and its free energy $F(\Delta_0)$
can not be expanded in $\Delta_0$.

 The relevant fields to enter the GL functional are:
$\Delta_0$, $\Delta_1$ and $S_z$.


 Assuming expansion in powers of small $S_z, \Delta_1$ near second
transition, the GL functional $F= F_{sc} + F_{magn} + F_{int}$ is:
\beqa
&F_{sc} = F(\Delta_0) + \alpha_1/2|\Delta_1|^2 + \alpha_2/4 |\Delta_1|^4 + 
\nonumber\\
&\beta |\nabla_i\Delta_1|^2, \alpha_{1,2}\geq 0\nonumber\\
&F_{magn} = {a_1(T)\over{2}}|S_z|^2 +  {a_2\over{4}}|S_z|^4 +
{a_3\over{2}}|\nabla_i S_z|^2   \nonumber\\
&F_{int} = -{b\over{2i}}(\Delta_0^*\Delta_1 - h.c.)S_z  
\label{int}
\eeqa
 $\Delta_0$ should enter in $F_{int}$ for it to be invariant under the
global $U(1)$ symmetry $\Delta_{0,1} \rightarrow \Delta_{0,1}
\exp(i\theta)$ \cite{heeb}.  Homogeneous solution will have lowest
energy and therefore gradient terms are take to be zero hereafter.

All but $F_{int}$ terms in the free energy Eq.(\ref{int}) are positive
and can not produce the instability of the original $d_{x^2-y^2}$
state.  $F_{int}$ can be negative since it is linear in $\Delta_1$ and
$S_z$ and this term is crucial in producing second transition.

Magnetic energy $F_{magn}$ has a temperature dependent coefficient 
\beqa
a_1(T) = a_1 n_{imp} (T-T_m)
\label{a1}
\eeqa
and vanishes at $T_m$, $a_1$ is dimensionless.  With this choice I
will assume that Ni impurities would order ferromagnetically at $T_m
\simeq 1K$ in the absence of the interaction with condensate.

Consider $F_{sc}$.  The second and third terms in $F_{sc}$ describe be
the energy cost of opening the fully gapped state with $\Delta_1$ when
interaction prefers to keep node, i.e.  pure $d_{x^2-y^2}$ state. The
change in free energy due to secondary order parameter is given by the
difference in energy of quasiparticles before and after $\Delta_1$
component is generated. One can calculate the change in the energy of
the superconductor subjected to the {\em homogeneous} external
$d_{xy}$ source field: $H_{xy}=
\kappa \sum_{\bk} \Delta_1(\bk) \psi^*_{\bk \sigma} \psi^*_{-\bk -\sigma}$, where $0 < \kappa < 1$ is the integration constant. Using standard result $\partial_{\kappa} F_{sc} = 1/\kappa \langle 
H_{xy} \rangle$ one finds increase of energy at $ \Delta_1 << T << \Delta_0$:
\beqa
&\delta F_{sc}=
{\alpha_1\over{2}}|\Delta_1|^2+{\alpha_2\over{4}}|\Delta_1|^4\nonumber\\
\label{Fsc}
\eeqa
Here $\alpha_1 \simeq N_0$, and $\Delta_1$ is taken to be constant on
the Fermi surface, see Eq.(\ref{F1int}), \cite{SNSBal,com1}.

Fix $\Delta_0$ to be real positive and the relative phase of $\Delta_1
= |\Delta_1|
\exp(i \nu)$ without loss of generality. Minimizing  the functional Eq.(\ref{int}) I find that the
 $\pi/2$ relative phase of $\Delta_1$ comes out naturally: phase will
be determined by minimization of energy:
\beqa
\nu = \pi/2 \ sgn(b S_z)
\eeqa
This choice takes the maximum advantage of the $L_zS_z$ coupling and
minimization {\em requires} a complex order parameter $\Delta_0 + i
\Delta_1$ in the low temperature phase. T and P are violated {\em
spontaneously} even with $T_m=0$.  The minimization yields:
\beqa
&S_z = {b\over{a_1 n_{imp}(T-T_m)}} \sin \nu \Delta_0|\Delta_1| \nonumber\\
&|\Delta_1|^2 = {1\over{\alpha_2}}({b^2\over{a_1 n_{imp}(T-T_m)}} \Delta^2_0 - \alpha_1)
 = \chi (T^*_c- T)\nonumber\\
&\delta F = -{\alpha_2\over{4}}|\Delta_1|^4  
\sim  \ |T-T^*_c|^2 \nonumber\\
&T^*_c \simeq T_m + 4\pi^2(N_0\Delta_0)^2 n_{imp}{g^2 N_0\over{a_1 a^2}}
\label{sol}
\eeqa
where I used the Eq.(\ref{b}) in the last line.  This is the main
result of this paper. Solution Eq.(\ref{sol}) indicates that the
transition is of the second order with the jump in the specific heat.
I assumed that $|\Delta_0(T)|^2/a_1(T-T_m)$ has a linear temperature
slope near $T^*_c$.  It follows from the solution Eq(\ref{sol}) that:

1) Even if $T_m = 0$ the ordering will occur at $T_c^* = 4\pi^2
n_{imp}(N_0\Delta_0)^2 {g^2 N_0\over{a_1 a^2}}$. However the softness
of the spin system near $T_m$ enhances the effect and makes $T^*_c
\geq T_m$ within this mean field approach. Taking the typical values
for Bi2212 of $\Delta_0 = 450 K, E_F = 1/N_0 = 3000 K$, assuming $a_1
\sim 1$ and taking the characteristic value of spin-rbit coupling of
Ni $g/a \sim 320 K$, I find $T_c^* \simeq 0.3 K$.  In the real system,
if the spin-glass freezing occurs, the freezing will occur at first
for the spins that are well separated. This will preclude the
Josephson tunneling between the patches, as discussed above. Only at
lower temperatures, when majority of spins are frozen, the tunneling
would be able to lock in the superconducting phase. Hence the real
phase ordering will occur at temperatures, substantially lower then
the mean field estimated $T_c^* \leq T_m$.

2) As the function of impurity concentration two effects occur
simultaneously. First, condensate density $|\Delta_0|^2$
decreases. Second, the suppression of $\Delta_1$ due to increased
impurity scattering will also lower $T_c^*$. These effects will lead
eventually to the disappearance of the transition. Quick suppression
of the transition temperature $T^*_c$ with impurity concentration
should be expected.

3) Strong magnetic field parallel to the layers,
 $ H \gg H_{c1,ab} \sim 1 \ Gauss$ in plane, will 
suppress the second phase.
In the field Ni spins will be aligned in the layers, 
linear coupling term on $H_{int}$ will
be zero and $d_{xy}$ component will vanish. This effect 
might explain the suppression
of the second transition by magnetic field 
$H_{c1,ab} \ll H \leq H_{c1,c} \sim 300 \  Gauss$, seen in experiment
\cite{Roman1}.

 Weak localization of the quasiparticles \cite{Lee}, in principle, can
cause the rapid decrease in the thermal conductivity.  Experimental
facts argue against localization in the layers for the following
reasons: the field {\em parallel} to the layers suppresses the
observed feature, it disappears at higher Ni concentration and the
specific heat is increased near transition temperature. Specific heat
and thermal conductivity in the field parallel to the layers will help
to determine how relevant localization of quasiparticles is to the
experimentally observed transition.

In the superconducting state with nonzero orbital current $L_z$ the
dominant fraction of the orbital moment is ``stored'' at the edge of
the sample, similar to $^3He-A$. The edge currents in $d+id$ state and
their topological characteristics was addressed recently in
\cite{Laugh1,Vol}.

To test the proposed state following experiments can be done. The
driving mechanism for the second phase clearly distinguishes between
magnetic and nonmagnetic impurities, hence more experiments on Bi2212
with nonmagnetic impurities will be helpful \cite{Roman1}. Theory
predicts the ordering of Ni moments below $T^*_c$, and one should be
able to detect magnetization in $\mu SR$ experiments or in ac
susceptibility. The increased superfluid density due to second
component translates into the change in the penetration depth below
$T^*_c$ which can be detected. These and other experiments will help
to resolve if the proposed mechanism is correct.

In conclusion, I presented the mechanism for a second order phase
transition of original d-wave state into $d+id$ state with
spontaneously broken T and P.  In the ordered phase both impurity
spins $S_z$ and $d_{xy}$ component of the order parameter develop and
are proportional to each other. The low temperature phase develops
magnetic moment both due to magnetic impurities spin and because of
the finite angular momentum of $d+id$ state.

I am grateful to R.B. Laughlin, D.H. Lee, A. Leggett, R. Movshovich,
  M. Salkola and G. Volovik for the useful discussions. This work was
  supported by US DOE.


\begin{references}

 

\bibitem{AG} A.A. Abrikosov and L.P. Gor'kov, Soviet Phys. JETP {\bf 12},1243 
(1961). P.W. Anderson, Phys. Rev. Lett. {\bf 3}, 325 (1959);
T. Tsuneto, Prog. Theor. Phys. {\bf 28}, 857 (1962);
D. Markowitz and L.P. Kadanoff, Phys. Rev. {\bf 131}, 563 (1963).


\bibitem{Roman1} R. Movshovich, et.al., cond-mat/9709061; private 
communication.

\bibitem{SNSBal} A.V. Balatsky, cond-mat/9709287, and in  Proceedings of Spectroscopies in Novel Superconductors, 1997, Journ. of Phys. Chem. of Solids, to be published.



\bibitem{Green1} M. Covington, et.al., Phys. Rev. Lett., {\bf 79}, 277, (1997).


\bibitem{Sauls1} M. Fogelstrom et.al., Phys. Rev. Lett., {\bf 79}, 281, (1997).

\bibitem{Laugh1} R.B. Laughlin, Preprint,  cond-mat/9709004. 

\bibitem{Ong1}  K. Krishana, et.al., Science, {\bf 277}, 83, (1997).

\bibitem{YSR} L. Yu, Physica Sinica {\bf 21}, 75, (1965); H.
 Shiba, Progr. Theor. Phys. {\bf 40}, 435, (1968); A.I. Rusinov,
Sov. Phys. JETP Letters {\bf 9}, 85, (1969).

\bibitem{Rama} T.V. Ramakrishnan, preprint.  Proceedings of Spectroscopies in Novel Superconductors, 1997, Journ. of Phys. Chem. of Solids, to be published.


\bibitem{Han} H. Hancotte, et. al.,  Phys. Rev. {\bf B 55}, R3410,(1997). 
In this paper authors argue about the substantial gap increase,
measured from zero energy to the maximum in the tunneling DOS, in
Bi2212 with 1.8\% Ni impurities, compared to the pure case. This
increase is consistent with the finite gap everywhere on the Fermi
surface due to $d_{xy}$ component generated by impurity
scattering. These authors also argue that the gap increase is observed
only for Ni doping and is {\em absent} in Zn doped samples. This fact
points to important gaping effect, which occurs only for magnetic
impurities.


\bibitem{SNSRoman} R. Movshovich, et.al, Proceedings of Spectroscopies in Novel Superconductors, 1997, Journ. of Phys. Chem. of Solids, to be published.  





\bibitem{SG} See for example D. Meschede et.al.,  Phys. Rev. Lett., {\bf 44}, 102, (1980); L. E. Wenger and P. H. Keesom, Phys. Rev., {\bf B 13}, 4053, (1976). Similar increase of the specific heat at low temperatures has been reported for Fe and Co -doped but not for Zn-doped
 Bi2212 in M.K. Yu and J. P. Franck, Phys. Rev., {\bf B 53}, 8651,
(1996).








\bibitem{heeb} This term was discussed in the context of the vortex solution 
 in an external magnetic field by R.Heeb, A. van Oterlo, M. Sigrist
and G. Blatter, Phys. Rev {\bf B 54}, 9385, (1996) and in M. Palumbo,
P. Muzikar and J. Sauls, Phys. Rev {\bf B 42}, 2681, (1990).  For a
general discussion of the GL functional in the presence of mixed
symmetry terms see also R. Joynt, Phys. Rev {\bf B 41}, 4271, (1990);
P.I. Soininen, C. Kallin and J. Berlinsky, Phys. Rev {\bf B 50},
13883, (1994); J. Xu, Y. Ren and C.S. Ting, Phys. Rev {\bf B 53},
R2991, (1996); H. Won and K. Maki, Phys. Rev {\bf B 53}, 5927, (1996).

  


\bibitem{com1}  In the opposite case of the  $\Delta_1$ component open only in the narrow vicinity of the node, within width $w \propto \Delta_1$ the angular 
integral in Eq.(\ref{Fsc}) is restricted to small angles near the
nodes.  I find in this case $F_{sc} \propto |\Delta_1|^3
\ln(E_c/max(\Delta_1, T))$ which is similar, apart from logarithmic
factor, to the result obtained by Laughlin \cite{Laugh1}.


\bibitem{Lee} P.A. Lee, Phys. Rev. Lett., {\bf 71}, 1887, (1993). See,
however, A. V. Balatsky and M. I. Salkola, Phys. Rev. Lett., {\bf 76},
2386, (1996).  \bibitem{Vol} G. E. Volovik, Pis'ma ZhETF {\bf 66},
492, (1997).  \end{references}
 \end{document}